\documentclass[10pt]{article}

\usepackage{amsmath}
\usepackage{amssymb}
\usepackage{graphics}
\usepackage{rotating}
\usepackage{cite}
\usepackage{color}
\usepackage{fancybox}
\usepackage{pstricks}
\usepackage{bm}
\usepackage{bbm}

\def\0{\mbox{\tiny $0$}}
\def\1{\mbox{\tiny $1$}}
\def\2{\mbox{\tiny $2$}}
\def\3{\mbox{\tiny $3$}}
\def\4{\mbox{\tiny $4$}}
\def\5{\mbox{\tiny $5$}}
\def\6{\mbox{\tiny $6$}}
\def\7{\mbox{\tiny $7$}}
\def\8{\mbox{\tiny $8$}}
\def\9{\mbox{\tiny $9$}}
\title{\shadowbox{\large \bf \begin{tabular}{c}
QUATERNIONIC DIRAC SCATTERING\end{tabular}}}

\author{
\normalsize \bf  Stefano De
Leo\thanks{Department of Applied Mathematics, State University of
Campinas, Brazil [deleo@ime.unicamp.br]}\,\,\,{\rm ,}\,\,\,\,\,Gisele Ducati\thanks{CMCC,
Universidade Federal do ABC, S\~ao Paulo, Brazil [ducati@ufabc.edu.br]}\,\,\,\,\,\,\,{\rm and}\,\,\,\,\,Sergio Giardino\thanks{Department of Applied Mathematics, State University of
Campinas, Brazil [giardino@ime.unicamp.br]}
}

\date{\small
\fcolorbox{black}{yellow} {\color{red} $\bullet$ {\color{black}{
{\footnotesize {\sc Journal of  Physical Mathematics} {\bf 6}, 1000130-6 (2015)
}}} {\color{red}{$\bullet$}} } }

\begin{document}
%

\maketitle

\vspace*{-.7cm}

\begin{abstract}
\noindent The scattering of a Dirac particle  has been studied for a quaternionic potential step. In the potential region an additional diffusion solution is obtained. The quaternionic solution which generalizes the complex one presents an amplification of the reflection and transmission rates.  A detailed analysis of the quaternionic spinorial  velocities sheed new light on the additional solution. For pure quaternionic potentials, the interesting and  surprising result of total transmission is found. This suggests that the presence of pure quaternionic  potentials cannot  be seen by analyzing the reflection or transmission rates. It has be observed  by  measuring the mean value of some operator.
\end{abstract}












\section*{\normalsize I. INTRODUCTION}

There are several proposals to formulate quantum mechanics\cite{Cohen,Zuber} by generalizing the complex number field. The most obvious generalization is done by introducing quaternions\cite{Adler:1995qqm,Q1,Q12,Q13,Q14}, but there are other possibilities, like Clifford algebras \cite{CA1} and $p-$adic numbers \cite{Brekke:1993gf}. Even inside each proposal, there may be internal divisions. In this article, we study a quaternionic quantum mechanics endowed with a quaternionic scalar product\cite{Adler:1995qqm}.

Our interest in the quaternionic formulation of Dirac equation  is motivated by the recent interesting results obtained in the standard Dirac theory for the  diffusion\cite{Dif1,Dif3}, tunneling\cite{Tun1} and Klein zones\cite{Klein1}. The generalization to quaternionic potentials\cite{Q14} could be useful to understand many hidden aspects of the Dirac solutions. Even simple solutions, as those involving the potential step, contain  interesting results. In the case of quaternionic quantum mechanics, there are  fewer studies which investigate the quaternionic formulation of Dirac equation. In  Adler's book \cite{Adler:1995qqm}, the theoretical foundations have been laid down, nevertheless explicit physical examples are still quite rare.
More recently, the Dirac solutions in the presence of quaternionic potentials have been detailed discussed in ref. \cite{Q14}. In the potential region, the quaternionic solutions allow two values for the wave-function  momentum. This represents the first important difference with respect to the complex case in which only one value is found.
A better understanding of the additional quaternionic solution can be achieved by studying an explicit example of quaternionic potential and calculating the reflection and transmission coefficients.  In order to do this, we recall the general solution presented in a previous article\cite{Q14} and study the scattering by a quaternionic potential step. In the complex scattering, for one-dimensional motion, we have not spin flip in the reflected and transmitted beams\cite{Zuber,Dif3}. Thus, the spin flip is not considered too for the quaternionic case. The additional quaternionic solutions  and the new reflection and transmission coefficients satisfy conservation of probability and in the complex limit reproduce the standard Dirac scattering. A novel and intriguing behavior is observed for the scattering by a purely quaternionic step. In this case, there is no reflection and the wave-function shows an oscillating spatial pattern similar to the one observed for circularly polarized light.

The article is organized as follows. In section II, for the convenience of the reader we fix our notation and recall the Dirac solution in the presence of a quaternionic potential. In section III, we obtain the reflection and transmission coefficients and calculate the probability current. In section IV, we discuss the effect of the quaternionic potential on the particle velocities. The last sections  round off the article with a brief discussion on the total transmission for pure quaternionic potentials (section V), final considerations and future possible investigations (section VI).

\section*{\normalsize II. THE QUATERNIONIC DIRAC EQUATION}

 As usual in quaternionic quantum mechanics\cite{Adler:1995qqm}, the Dirac equation is written in terms of an anti-hermitian Hamiltonian operator. In this spirit, we introduce the anti-hermitian quaternionic  potential step
\[\left\{\,0\,\,\,\mbox{for}\,z<0\,\,\,(\mbox{\sc region I})\,\,\,,\,\,\,\,\,
i\,V_{\0}+j\,V_{\1}+k\,V_{\2}\,\,\,\mbox{for}\,z>0\,\,\,(\mbox{\sc region II})\,
\right\}\]
where $V_{\0,\1,\2}$ are real constants. In the potential region (region II), the Dirac equation then reads
\begin{equation}\label{dirac1}
\partial_t\Psi(\boldsymbol{r},t) = -
\left[\, \boldsymbol{\alpha} \cdot \nabla
+i\,m\,\beta+i\,V_{\0}+j\,V_{\1}+k\,V_{\2} \, \right]\,\Psi(\boldsymbol{r},t).
\end{equation}
 For $V_{\1}=V_{\2}=0$, the complex step is recovered. The Dirac Hamiltonian conatins the matrices $\bm{\alpha}$ and $\beta$ which satisfy the algebra
\begin{equation}
\bm{\alpha}=\bm{\alpha}^\dagger\,,\qquad\beta=\beta^\dagger\,,
\qquad\bm{\alpha}^2=\beta^2=1\,,
\qquad\{\,\beta\,,\,\bm{\alpha}\,\}=0\qquad\mbox{and}\qquad\{\alpha_r,\,\alpha_r\}=2\,\delta_{rs}\,\mathbbm{1}.
\end{equation}
In the Dirac representation\cite{Zuber}, we have
\begin{equation}
\bm{\alpha}=\left(
\begin{array}{cc}
0&\bm{\sigma}\\
\bm{\sigma}&0\\
\end{array}\right),\qquad
\beta=\left(
\begin{array}{cc}
\mathbbm{1}&0\\
0&-\mathbbm{1}\\
\end{array}
\right),
\end{equation}
where $\bm \sigma$ are the well known complex Pauli matrices. For convenience of notation, the quaternionic potential
can be rewritten as follows
\[W_{\0} =V_{\2}+i\,V_{\1} = |W_{\0}|\,\exp\big[\,i\,\arctan(V_{\1}/V_{\2})\,\big]=  |W_{\0}|\,\exp\big[\,i\,\phi\,\big]\,\,. \]
Considering  a motion along the $z$-axis, $\boldsymbol{p} = (\,0\,,\,0\,,\,Q\,)$, introducing the
the quaternionic spinor
\begin{equation}
\label{qwf}
\Psi(\boldsymbol{r},t)=\big[\,u\,+\,j\,w\,\big]\,\exp[\,i\,(\,Q\,z\,-E\,t\,)\,]\,\,,
\end{equation}
where $u$ and $v$ are complex spinors, we obtain two coupled complex  matrix equations,
\begin{eqnarray}
\label{s2}
\left(\,E-Q\,\alpha_{\3}-m\,\beta\,-V_{\0}\right)\,u & = & -\,W_{\0}^{^{*}}w  \nonumber \\
\left(\,E-Q\,\alpha_{\3}+m\,\beta\,+V_{\0}\right)\,w & = & -\,W_{\0}\,u\,\,.
\end{eqnarray}
The first equation is obtained from the  complex part of equation (\ref{dirac1}), and the second one   is obtained
taking  the pure quaternionic part. Equations (\ref{s2}) can be rewritten as eigenvalue equations,
\begin{eqnarray}
\left(\,E-Q\,\alpha_{\3}+m\,\beta\,+V_{\0}\right)\,\left(\,E-Q\,\alpha_{\3}-m\,\beta\,-V_{\0}\right)\,u\, & = & \left|W_{\0}\right|^{^{2}} u\,\,,\nonumber \\
\left(\,E-Q\,\alpha_{\3}-m\,\beta\,-V_{\0}\right) \left(\,E-Q\,\alpha_{\3}+m\,\beta\,+V_{\0}\right)\,w\, & = & \left|W_{\0}\right|^{^{2}} w\,\,.\label{matrix2}
\end{eqnarray}
The determinants of the matrices on the left hand side of equations (\ref{matrix2}) are equal, and consequently the eigenvalues obtained from any of them are also equal. Non-trivial solutions have to  satisfy
\begin{equation}\label{det}
 \mbox{det}\left[\,\left(\,E-Q\,\alpha_{\3}+m\,\beta\,+V_{\0}\right)\,
\left(\,E-Q\,\alpha_{\3}-m\,\beta\,-V_{\0}\right) - \left|W_{\0}\right|^{^{2}} \,\right]=0.
\end{equation}
From equation (\ref{det}), we obtain  two squared momenta for particles moving in the region II, namely
\begin{equation}\label{momenta}
Q^{^{2}}_{_{\pm}}=q^{\2}_{_{\pm}}+\,\left|W_{\0}\right|^{^{2}} \pm \,2\delta\,\,,
\end{equation}
where
\begin{equation*}
q^{\2}_{_{\pm}} = (E\pm V_{\0})^{^{2}} -m^{\2},\qquad \delta=\sqrt{(\,E\,V_{\0}\,)^{^{2}}+ (\,p\,|W_{\0}|\,)^{^{2}}} -E\,V_{\0}\qquad\mbox{and}\qquad p=\sqrt{E^2-m^2}\,\,.
\end{equation*}
From equation (\ref{momenta}), $Q_{_{+}}^{^{2}}>0$, and $Q_{_{-}}^{^{2}}$ may be either positive or negative. Hence, we find  three energy zones determined  by the $Q_{_{-}}^{^{2}}$ value,
\begin{eqnarray}
&1)& \mbox{diffusion zone}\qquad \;Q_{_{-}}^{^{2}} >0\qquad\mbox{and}\qquad E>\sqrt{|W_{\0}|^{^{2}}+(V_{\0}+m)^{^{2}}}\,\,,\nonumber\\
&2)&\mbox{tunneling zone}\qquad Q_{_{-}}^{^{2}}<0\qquad\mbox{and}\qquad\text{max}\Big[m,\,\sqrt{|W_{\0}|^{^{2}}+(V_{\0}-m)^{^{2}}}\,\Big]<E<
\sqrt{|W_{\0}|^{^{2}}+(V_{\0}+m)^{^{2}}}\,\,,\nonumber\\
&3)& \mbox{Klein zone}\;\;\;\;\;\;\qquad Q_{_{-}}^{^{2}} >0\qquad\mbox{and}\qquad E<\sqrt{|W_{\0}|^{^{2}}+(V_{\0}-m)^{^{2}}}\,\,.\nonumber
\end{eqnarray}
These regions generalize the diffusion, tunneling, and Klein zones obtained for the complex case\cite{Dif1,Dif3,Tun1,Klein1},
 \begin{eqnarray}
&1)& \mbox{diffusion zone}\qquad \;\;q_{_{-}}^{^{2}} >0\qquad\mbox{and}\qquad E>V_{\0}+m\,\,,\nonumber\\
&2)&\mbox{tunneling zone}\qquad q_{_{-}}^{^{2}}<0\qquad\mbox{and}\qquad V_{\0}-m<E< V_{\0}+m\,\,,\nonumber\\
&3)& \mbox{Klein zone}\;\;\;\;\;\;\;\qquad q_{_{-}}^{^{2}} >0\qquad\mbox{and}\qquad E< V_{\0}-m\,\,.\nonumber
\end{eqnarray}
For diffusion\cite{Dif1,Dif3} and Klein\cite{Klein1} zones, we find oscillatory solutions interpreted as particles or anti-particles. In the tunneling zone\cite{Tun1}  the wave function becomes evanescent and thus we have not particle or anti-particle propagation.

In order to explicitly determine the quaternionic wave functions, we calculate the eigenvectors of equations (\ref{matrix2}). The four component eigenvectors can be conveniently written by defining the two-dimensional vectors $\boldsymbol\chi=\{(1,\,0)^t,\,(0,\,1)^t\}$. After simple algebraic manipulation,  for $Q=Q_{_-}$, we find
\begin{equation}\label{wfminus}
u = \left[\,
\begin{array}{c}
 \boldsymbol{\chi} \\  \\ \displaystyle{\frac{Q_{_{-}}}{E-V_{\0}+m -\frac{\delta}{E-m}}}\,\,\sigma_{\3}\,\boldsymbol{\chi}
 \end{array}
 \right]\qquad\,\,\,\,\,\,\,\mbox{and}\,\,\,\,\,\,\, w\, = - \,W_{\0}\, \left(\,E-Q_{_{-}}\,\alpha_{\3}+m\,\beta\,+V_{\0}\right)^{^{-1}}u\,\,,
\end{equation}
and  for $Q=Q_{_+}$,
\begin{equation}\label{wfplus}
\widetilde w  = \left[\,
\begin{array}{c}
  \displaystyle{\frac{Q_{_{+}}}{E+V_{\0}+m +\frac{\delta}{E-m}}}\,\,\sigma_{\3}\,\boldsymbol{\chi} \\ \\
  \boldsymbol{\chi}
 \end{array}
 \right]\qquad \,\,\,\,\,\,\mbox{and}\,\,\,\,\,\,\,
\widetilde{u} = - \,W^{^{*}}_{\0}\, \left(\,E-Q_{_{+}}\,\alpha_{\3}-m\,\beta\,-V_{\0}\right)^{^{-1}}\widetilde{w}\,\,.
\end{equation}

\section*{\normalsize III. THE MATCHING CONDITIONS}
In order to study the scattering of a Dirac particle by the quaternionic step,
for simplicity of calculation we choose a spin orientation for the incoming particle, i.e.
$\boldsymbol{\chi}=(1\,\,0)^{t}$ and explicit in the complex spinors $u$, $w$, $\widetilde{u}$ and
$\widetilde{w}$ their dependence  on $Q$, $V_{\0}$ and $W_{\0}$. For example, in region I, we have an incident  particle with momentum $p=\sqrt{E^2-m^2}$, thus represented by $u[Q=p\,;\,V_{\0}=0\,,\,W_{\0}=0\,]$. Obviously, in the same region there are reflected particles with momenta $-p$, thus represented by $u[-p\,;\,0\,,\,0\,]$ . There is no spin flip in the reflection or transmission, thus we can eliminate the zero component in our wave functions and work with two-component quaternionic spinors.

In region I (potential free region), the quaternionic wave function is then given by
\begin{equation}
\Psi_{_I}(z)=u[p;\,0,\,0]\,\exp[i\,p\,z] + \Big[\,u[-p;\,0,\,0]\,R + j\,\widetilde{w}[-p;\,0,\,0]\,\widetilde{R}\,\Big]\,\exp[-\,i\,p\,z]\,\,,
\end{equation}
 where $R$ and $\widetilde R$ are complex coefficients. Note that the presence of complex exponential in (\ref{qwf}) requires multiplication from the right by complex coefficients. In region II (potential region), the quaternionic wave function is characterized by two possible momenta, $Q_{_+}$ and $Q_{_-}$, and its explicit expression is
\begin{eqnarray}
 \Psi_{_{II}}(z)&=&\Big[\,u[Q_{_{-}};\,V_{\0},\,W_{\0}] + j\,\,w[Q_{_{-}};\,V_{\0},\,W_{\0}]\,\Big]\,T \, \exp[i\,Q_{_{-}}\,z] \,+\nonumber\\
& & \Big[\,\widetilde{u}[Q_{_{+}};\,V_{\0},\,W_{\0}] + j\,\widetilde{w}[Q_{_{+}};\,V_{\0},\,W_{\0}]\,\Big]\,\widetilde{T}\, \exp[i\,Q_{_{+}}\,z]\,\,.
\end{eqnarray}
As happens for the reflection coefficients, the transmission coefficients $T$ and $\widetilde T$ are also complex.
The matching condition, $\Psi_{_I}(0)=\Psi_{_{II}}(0)$, implies
\begin{eqnarray}
u[p;\,0,\,0] + u[-p;\,0,\,0]\,\,\,R & = & u[Q_{_{-}};\,V_{\0},\,W_{\0}]\,\,\,T +  \widetilde{u}[Q_{_{+}};\,V_{\0},\,W_{\0}]\,\,\,\widetilde{T}\nonumber \\
\widetilde{w}[-p;\,0,\,0]\,\,\,\widetilde{R}  & =  &  w[Q_{_{-}};\,V_{\0},\,W_{\0}]\,\,\,T +   \widetilde{w}[Q_{_{+}};\,V_{\0},\,W_{\0}]\,\,\,\widetilde{T}\,\,.
\end{eqnarray}
From these coupled equations, we can obtain the complex coefficients $R$, $\widetilde R$, $T$, $\widetilde T$ in terms of the wave function parameters and consequently calculate the transmission and reflection rates.

\subsection*{\normalsize III.A THE REFLECTION AND TRANSMISSION COEFFICIENTS}
To simplify our notation let us introduce the adimensional quantity
\[ a=\frac{p}{E+m}\]
and explicitly rewrite the quaternionic wave function in region I,
\begin{equation*}
 \Psi_{_{I}}(z) = \left(\begin{array}{c} 1 \\ a \end{array} \right) \exp[\,i\,p\,z\,] +\left(\begin{array}{c} 1 \\ -\,a \end{array} \right)\,R\, \exp[\,-\,i\,p\,z\,] + j\,\left(\begin{array}{c} -\,a \\ 1 \end{array} \right) \,\widetilde{R}\, \exp[\,-\,i\,p\,z\,]\,\,,
 \end{equation*}
and region II,
\begin{equation*}
\Psi_{_{II}}(z) = \left[\, \left(\begin{array}{c} 1 \\ A_{_{-}} \end{array} \right) - j\,W_{\0}\, \left( \begin{array}{c} M_{_{-}} \\ N_{_{-}} \end{array} \right)\,\right]\,T\,\exp[\,i\,Q_{_{-}}z\,]\,\,
+
 \left[\,- W_{\0}^{^{*}}\, \left( \begin{array}{c} N_{_{+}} \\ M_{_{+}} \end{array} \right)+j\,  \left(\begin{array}{c} A_{_{+}}\\1  \end{array} \right) \,\right]\,\widetilde{T}\,\exp[\,i\,Q_{_{+}}z]\,\,,
\end{equation*}
where
\[
A_{_\pm}= \frac{Q_{_\pm}}{E\pm V_{\0}+m\pm \frac{\delta}{E-m}}\,\,,\,\,\,\,\,
M_{_\pm}= \frac{Q_{_\pm}A_{_\pm} + E-m \mp V_{\0} }{q_{_\mp}^{\2} - Q_{_\pm}^{^{2}}}\,\,,\,\,\,\,\,
N_{_\pm}= \frac{ (E+m \mp V_{\0})A_{_\pm}+ Q_{_\pm} }{q_{_\mp}^{\2} - Q_{_\pm}^{^{2}}}\,\,.
\]
The continuity equations at $z=0$ lead to the following system
\begin{eqnarray}
\left(\begin{array}{c} 1 \\ a \end{array} \right) + \left(\begin{array}{c} 1 \\ - a \end{array}\right)\,R & = & \left(\begin{array}{c} 1 \\ A_{_{-}} \end{array} \right) \,T \, -\,W^{^{*}}_{\0} \left( \begin{array}{c} N_{_{+}} \\ M_{_{+}} \end{array} \right)  \widetilde{T}\,\,,\nonumber \\
\left(\begin{array}{c}
-a \\ 1 \end{array}  \right)\,\widetilde{R} & = &  - \,W_{\0} \left( \begin{array}{c} M_{_{-}} \\ N_{_{-}} \end{array} \right) \,T\, +  \left(\begin{array}{c} A_{_{+}} \\ 1 \end{array} \right) \,\widetilde{T}\,\,.
\end{eqnarray}
After algebraic manipulations, we find
\begin{eqnarray}
\label{coef}
R &=& \frac{(a-A_{_{-}})(a+A_{_{+}}) + |W_{\0}|^{^{2}} (M_{_{+}} - a\,N_{_{+}})\,(M_{_{-}} +a\, N_{_{-}})}
{(a+A_{_{-}})(a+A_{_{+}}) - |W_{\0}|^{^{2}} (M_{_{+}} + a\,N_{_{+}})\,(M_{_{-}} +a\, N_{_{-}})}\,\,,\nonumber \\
T&=&\frac{2\,a\,(a+A_{_{+}})}{(a+A_{_{-}})(a+A_{_{+}}) - |W_{\0}|^{^{2}} (M_{_{+}} + a\,N_{_{+}})\,(M_{_{-}} +a\, N_{_{-}})}\,\,,\nonumber \\
\widetilde{R} &=& W_{\0}\,\frac{M_{_{-}} - A_{_{+}} N_{_{-}}}{a+A_{_{+}}}\,\,T\,\,,\nonumber \\
\widetilde{T} &=& W_{\0}\,\frac{M_{_{-}} +a N_{_{-}}}{a+A_{_{+}}}\,\,T\,\,.
\end{eqnarray}
In the complex limit, $W_{\0}\to 0$, we recover the standard reflection and transmission rates\cite{Zuber},
\begin{equation}
\widetilde{R}_c=\widetilde{T}_c=0\,\,,
\end{equation}
and
\begin{eqnarray}
R_c & =&  \frac{p\,(E-V_{\0}+m) -q_{_{-}}\,(E+m)}{p\,(E-V_{\0}+m) +q_{_{-}}\,(E+m)}\,\,,\nonumber\\
T_c &=&\frac{2\,p\,(E-V_{\0}+m)}{p\,(E-V_{\0}+m) +q_{_{-}}\,(E+m)}\,\,.
\end{eqnarray}

\subsection*{\normalsize III.B THE CURRENT PROBABILITY DENSITY}
The relation between the reflection and transmission coefficients can be obtained by using the continuity equation
\begin{equation}\label{continuity}
\frac{\partial\rho}{\partial t}+\bm{\nabla\cdot J}=0\,\,,
\end{equation}
where
\[ \rho=\Psi^\dagger\Psi\qquad\mbox{and}\qquad \bm{J}=\Psi^\dagger\bm{\alpha}\,\Psi\,\,.  \]
Observing that $\Psi^{\dagger}\Psi$ is independent of time  and the spatial dependence of $\Psi$ is only on the $z$-coordinate, we find
\[ \partial_{z}\, \Psi^\dagger\alpha_{\3}\,\Psi=0\,\,.    \]
The continuity of the current density between the free region I and the potential region II implies
\[\Psi_{_{\rm inc}}^{^{+}}\,\alpha_{\3}\,\Psi_{_{\rm inc}} + \Psi_{_{R}}^{^{+}}\,\alpha_{\3}\,\Psi_{_{R}} + \Psi_{_{\widetilde{R}}}^{^{+}}\,\alpha_{\3}\,\Psi_{_{\widetilde{R}}} =    \Psi_{_{T}}^{^{+}}\,\alpha_{\3}\,\Psi_{_{T}}  +  \Psi_{_{\widetilde{T}}}^{^{+}}\,\alpha_{\3}\,\Psi_{_{\widetilde{T}}}. \]
For the previous equation, we find
\begin{equation}\label{flux}
\displaystyle{|R|^{^{2}}+  |\widetilde{R}|^{^{2}} + \underbrace{\left(\, \frac{A_{_{-}} +|W_{\0}|^{^{2}} M_{_{-}}^{^{*}}N_{_{-}}}{2\,a} + \mbox{h.c.}\,\right)}_{\mbox{ $\rho$}}\,|T|^{^{2}} + \underbrace{\left(\, \frac{A_{_{+}} +|W_{\0}|^{^{2}} M_{_{+}}^{^{*}}N_{_{+}}}{2\,a} + \mbox{h.c.}\,\right)}_{_{\mbox{ $\widetilde{\rho}$}}}\,|\widetilde{T}|^{^{2}}  } =1\,\,.
\end{equation}
In the complex limit, $W_{\0}\to 0$, we recover the well know continuity equation
\begin{equation}
\label{contc}
|R_c|^{^{2}} + \underbrace{\frac{(q_{_-} +\,q_{_-}^*)\,(E+m)}{2\,p\,(E-V_{\0}+m)}}_{\mbox{ $\rho_c$}}\,\,|T_c|^{^{2}}=1\,\,.
\end{equation}
For $E>V_{\0}+m$, we have $q_{_-}\in \mathbb{R}$ and $\rho_c>0$.  Eq.(\ref{contc}) then implies a reflection rate $<1$ (diffusion zone). For $V_{\0}-m<E<V_{\0}+m$, we have  $q_{_-}^{^{2}}<0$. Consequently $\rho_c=0$ and  Eq.(\ref{contc}) implies a reflection rate $=1$. Finally, for $E<V_{\0}-m$, we have $q_{_-}\in \mathbb{R}$ and $\rho_c<0$, we find a reflection rate $>1$ suggesting pair production\cite{Zuber,Klein1}. In Fig.\,1, we plot the reflection rates for different value of quaternionic potentials as a function of $E/m$. In the Klein zone, the reflection rate increases by increasing the quaternionic potential.  In this energy zone, the quaternionic perturbation thus contributes to the phenomenon of pair production. This is confirmed by the transmission rates in Fig.\,2. It is important to observe that for complex tunneling only evanescent waves, $\exp[-|q_{_-}|z]$, appears in the potential region and this is seen by observing (Fig.1a) that the reflection rate is 1. The presence of a quaternionic potential breaks total reflection and this is explained by the fact that  together evanescent wave, $\exp[-|Q_{_-}|z]$, also   oscillatory waves, $\exp[i|Q_{_+}|z]$, are present in the potential region. The numerical calculation for the combined reflection and transmission coefficients is shown in Fig.\,3 and confirms the continuity equation (\ref{flux}).
From Fig.\,3, we also observe a shift in the starting point of the diffusion zone and more important we have a decreasing energy zone for the evanescent waves by increasing the quaternionic potential. The last observation clearly suggest that oscillatory waves kill evanescent waves in presence of great quaternionic potentials. Finally, we also find that the reflection and transmission rates $|\widetilde{R}|$ and $|\widetilde{T}|$ are very small compared to the reflection rates $|R|$ and $|T|$, thus as a first approximation we can use these last coefficient rates to study quaternionic perturbations of complex quantum scattering problems.

\section*{\normalsize IV. GROUP VELOCITIES ANALYSIS}

The incoming complex plane wave has the standard oscillatory exponential, $\exp[\,i\,(p\,z-E\,t]$, and consequently its group velocity is given by
\begin{equation}
v_{_{\rm in}} = \frac{\mbox{d}E}{\mbox{d}p} = \frac{p}{E}\,\,.
\end{equation}
The transmitted quaternionic plane waves contain, in the diffusion and Klein zones, the oscillatory exponentials
\[
\exp\left[\,i\,\left(\,Q_{_{\pm}}z-E\,t\,\right)\,\right]\,\,,
\]
which determine the following group velocities
\[
v_{_{\pm}}= \frac{\mbox{d}E}{\,\,\,\mbox{d}Q_{_{\pm}}}\,\,.
\]
Recalling that
\[Q_{_{\pm}}^{^{2}} = (E\pm V_{\0})^{^{2}} - m^{\2} +|W_{\0}|^{^{2}} \pm 2\,\left(\,\sqrt{E^{^{2}}(V_{\0}^{^{2}}+|W_{\0}|^{^{2}}) - m^{\2}|W_{\0}|^{^{2}}} -EV_{\0}\,\right)\,\,, \]
we immediately obtain
\[ Q_{_{\pm}} \mbox{d}Q_{_{\pm}} =\left[\, E\pm V_{\0} \pm \frac{E\,(V_{\0}^{^{2}}+|W_{\0}|^{^{2}})}{\sqrt{E^{^{2}}V_{\0}^{^{2}}+p^{\2}|W_{\0}|^{^{2}}}} \, \mp V_{\0}\,\right]\,\mbox{d}E\,\,,    \]
and consequently
\begin{equation}
v_{_{\pm}}= \frac{\,\,Q_{_{\pm}}}{E}\,\left(\,1 \pm \frac{V_{\0}^{^{2}}+|W_{\0}|^{^{2}}}{\sqrt{E^{^{2}}V_{\0}^{^{2}}+p^{\2}|W_{\0}|^{^{2}}}}\,\right)^{^{-1}}\,\,.
\end{equation}
The additional quaternionic spinor is then characterized by a velocity $v_{_{+}}$ which is a typical velocity of diffusion, see Fig.\,4a. Indeed, we have not tunneling or Klein zones as happens for the velocity $v_{_{-}}$, see Fig.\,4b.  By increasing the  quaternionic part of the potential we see decreasing the velocity of the additional quaternionic spinor and by increasing the incoming energy we obtain an increasing value of the velocity. These are typical effects of diffusion.  As observed before, the velocity $v_{_{-}}$ presents three different energy zones (see Fig.\,4b),  the diffusion zone for $E>\sqrt{|W|_{\0}^{^{2}}+\,(V_{\0}+m)}$ where the potential acts decelerating the particles, the Klein zone  for $E<\sqrt{|W|_{\0}^{^{2}}+\,(V_{\0}-m)}$ where the negative velocities solutions generates a flux from the right to the left which gives a reflection rates greater than one and consequently interpreted in terms of pair production and finally the intermediate  tunneling zone where we have not propagation
due to the presence of evanescent solutions. It is interesting to observe that in the diffusion zone, we have two propagation regimes and consequently two different velocities. These different velocities permit to decouple the two quaternionic solutions. In the pure quaternionic  limit ($V_{\0}\to 0$), as we shall discuss in our conclusions, these velocities tend to the same value and this value is $p/E$ the free propagation velocity.

\section*{\normalsize V. TOTAL TRANSMISSION}

Our analysis  shows a  surprising effect for pure quaternionic potentials, $V_{\0}=0$. In this case,
\[ Q_{_{\pm}}= p \,\pm\, |W_{\0}|,\qquad
A_{_{\pm}}=a,\qquad
M_{_{\pm}}=\mp\,\frac{a}{|W_{\0}|},\qquad\mbox{and}\qquad
N_{_{\pm}}=\mp\frac{1}{|W_{\0}|}\,\,,
\]
and the continuity equation  leads to the following system for the reflection and transmission coefficients,
\begin{equation}
1\pm R_q=T_q+\widetilde T_q\,e^{-i\phi}\qquad\mbox{and}\qquad \pm \widetilde R_q= -\,T_q\,e^{i\phi}+\widetilde T_q\,\,.
\end{equation}
The solutions are immediately obtained by showing total transmission,
\begin{equation}
R_q=\widetilde R_q=0\qquad T_q=\frac{1}{2}\qquad\mbox{and}\qquad \widetilde T_q=T_qe^{i\phi}.
\end{equation}
The transmitted wave-function for the pure quaternionic potential is then given  by
\begin{eqnarray}
\Psi_{_{q,T}} & = &
\left[\,\left(\begin{array}{c}
1\\a\\
\end{array}\right)\,\cos(|W_0|z)
-k\left(\begin{array}{c}
a\\1\\
\end{array}\right)\,e^{i\phi}\,\sin(|W_0|z)\,\right]\,e^{ipz}\nonumber \\
 & = & \left[\,
\begin{array}{rr}
\cos(|W_0|z) &\,\,\,-k\, e^{i\phi}\, \sin(|W_0|z)\\
-k\, e^{i\phi}\,\sin(|W_0|z)&\cos(|W_0|z)
\end{array}\,\right]\,\,\underbrace{\left(\begin{array}{c}
1\\a\\
\end{array}\right)\,e^{ipz}}_{\Psi_{_{c,{\rm inc}}}}\,\,,
\end{eqnarray}
and propagates with group velocity $p/E$.  Due to the fact that the transmitted quaternionic spinor satisfies
\[ \Psi_{_{q,T}}^{^{\dag}}\Psi_{_{q,T}}=1\,\,,  \]
and moves with the same velocity of the free incoming particle {\em seems} impossible to see the presence of pure quaternionic potentials.  Nevertheless the problem of an apparent invisible presence of pure quaternionic potential can be solved by taking the mean value of operators. For example, the $z$ component spin  mean value for the
complex wave function $\Psi_{_{c,{\rm inc}}}$ is given by
\begin{equation}
\langle\,S_{_{z}}\,\rangle_{_c} = \frac{\Psi_{_{c,{\rm inc}}}^{^{\dag}}\,S_{_{z}}\,\Psi_{_{c,{\rm inc}}}}{\Psi_{_{c,{\rm inc}}}^{^{\dag}}\Psi_{_{c,{\rm inc}}}}=\frac{1}{2}\,\frac{\left(\,1\,\,\,a\,\right)\,\left(\begin{array}{rr}
1&0\\ 0 &-1 \end{array}\right)\,\left(\begin{array}{c} 1\\ a \end{array}\right)}{1+a^{\2}}
= \frac{1-a^{\2}}{2\,(1+a^{\2})}= \frac{m}{2\,E}\,\,,
\end{equation}
 which in the non-relativistic limit, $p<<m$, reproduces the well-known result $\langle\,S_{_{z}}\,\rangle_{_c} =1/2$.
 In the presence of a pure quaternionic potential, we should find
 \begin{eqnarray}
\langle\,S_{_{z}}\,\rangle_{_q} &=& \frac{\Psi_{_{q,{\rm inc}}}^{^{\dag}}\,S_{_{z}}\,\Psi_{_{q,{\rm inc}}}}{\Psi_{_{c,{\rm inc}}}^{^{\dag}}\Psi_{_{c,{\rm inc}}}}=\frac{1}{2}\,\frac{\left(\,1\,\,\,a\,\right)\,\left[\begin{array}{rr}
\cos (2\,|W_{\0}|\,z)&-\,k\,e^{i\phi}\,\sin(2\,|W_{\0}|\,z)\\
 k\,e^{i\phi}\,\sin(2\,|W_{\0}|\,z) & - \cos (2\,|W_{\0}|\,z)
 \end{array}\right]\,\left(\begin{array}{c} 1\\ a \end{array}\right)}{1+a^{\2}} \nonumber \\
  & = &  \frac{1-a^{\2}}{2\,(1+a^{\2})} \,\,\cos (2\,|W_{\0}|\,z) = \frac{m}{2\,E}\,\,\cos (2\,|W_{\0}|\,z)\,\,,
\end{eqnarray}
 which implies a $z$-dependence on the mean value of spin similar to what happens for circular polarized light\cite{Wolf}.

\section*{\normalsize VI. CONCLUSIONS}
In this article we have studied the scattering of a Dirac particle by a quaternionic potential step. Our analysis shows the presence of an additional quaternionic solution characterized by a group diffusion velocity $v_{_{+}}$. The quaternionic solution which generalizes the standard one is characterized by a group velocity $v_{_{-}}$ and presents three energy zones, diffusion-tunneling-Klein zones. For diffusion the two solution are, in general, different and could be decouple the two solution in the potential region showing the presence of a quaternionic potential. By increasing the  quaternionic part of the potantial we increase  the Klein zone and decrease the tunneling zone, see Fig.\,4b.

The most interesting effect is seen for a pure quaternionic potential where total transmission and
a $z$ dependence on the spin operators appear. This suggests a possible investigation in quaternionic optics\cite{Q13}.  Before to do it, due to the fact that we have to introduce Gaussian laser,
 in a forthcoming, we aim to extend the plane wave analysis done in this article to quaternionic wave packets\cite{Q12}.


%
%
%
%


\newpage

\begin{figure}
\vspace*{-3cm} \hspace*{-2cm}
\includegraphics[width=19.5cm, height=26cm, angle=0]{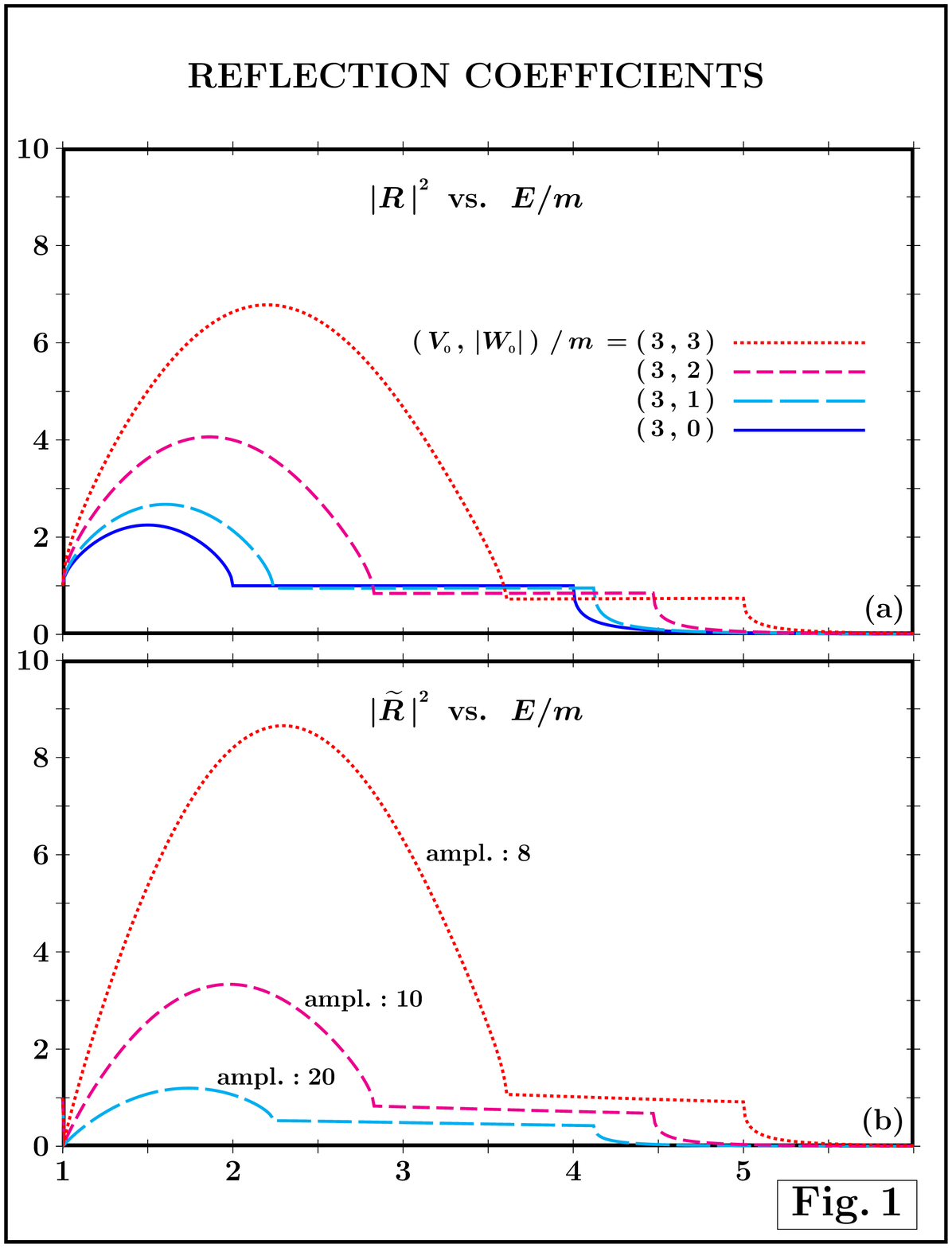}
\vspace*{-2.5cm}
 \caption{The reflection rates as a function of the incoming energy for a fixed complex potential $V_{\0}=3\,m$ (blue continuous line) and  for quaternionic perturbation $|W_{\0}| = (1,2,3)\, m$  (cyan/magenta/red dashed and dotted lines). By increasing the quaternionic part of the potential, the tunneling zone decreases. We also observe that in presence of quaternionic perturbation, due to the  additional diffusion solution, we have not total reflcetion in the tunneling zone. }
\end{figure}

\newpage

\begin{figure}
\vspace*{-3cm} \hspace*{-2cm}
\includegraphics[width=19.5cm, height=26cm, angle=0]{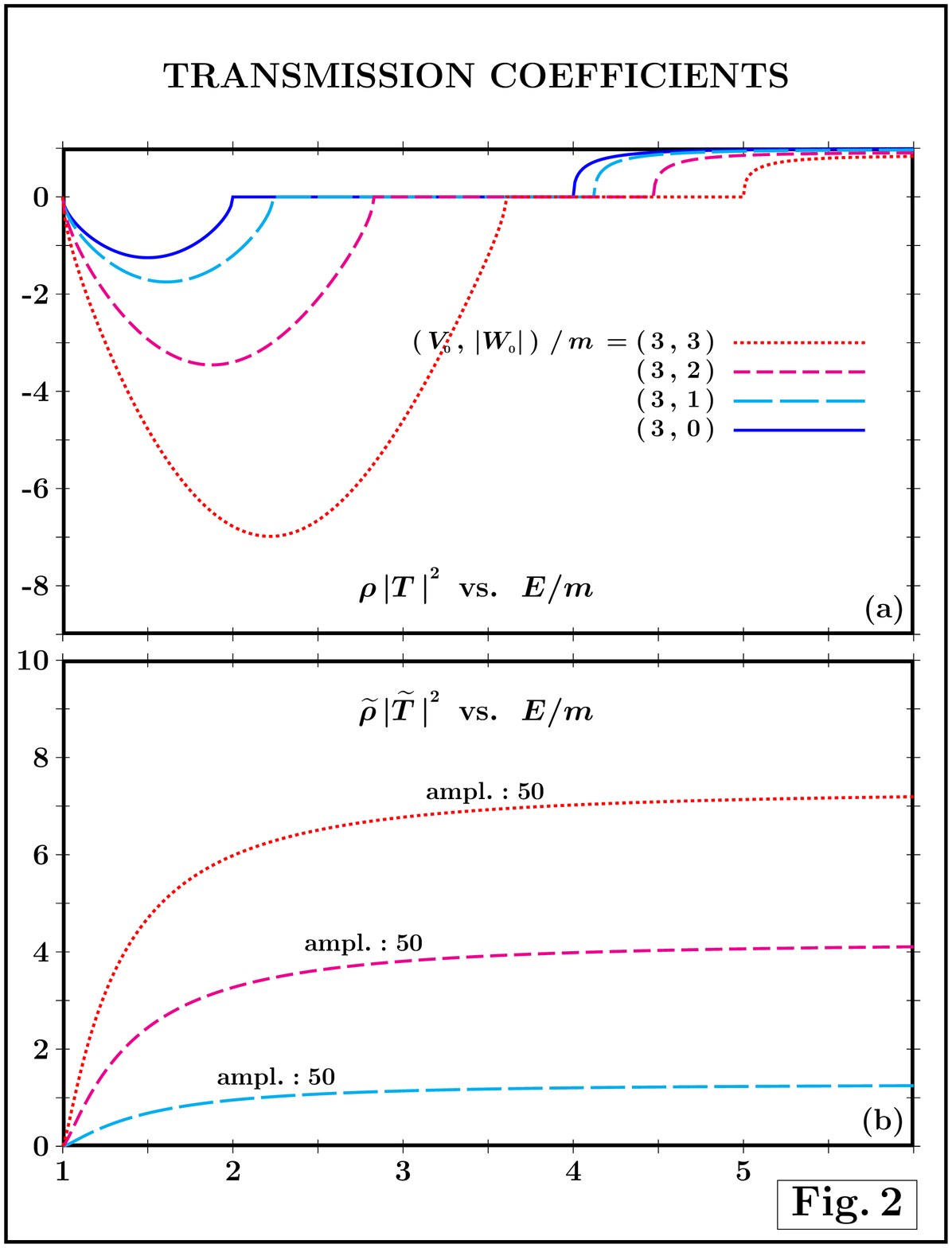}
\vspace*{-2.5cm}
 \caption{The transmission rates are plotted as function of the incoming energy for a fixed complex potential and different quaternionic perturbation. As observed in the caption of Fig.\,1, in presence of quaternionic potentials   total reflection breaks down. This is clear from the plot in (b) which confirms the presence of additional diffusion solutions.     }
\end{figure}

\newpage

\begin{figure}
\vspace*{-3cm} \hspace*{-2cm}
\includegraphics[width=19.5cm, height=26cm, angle=0]{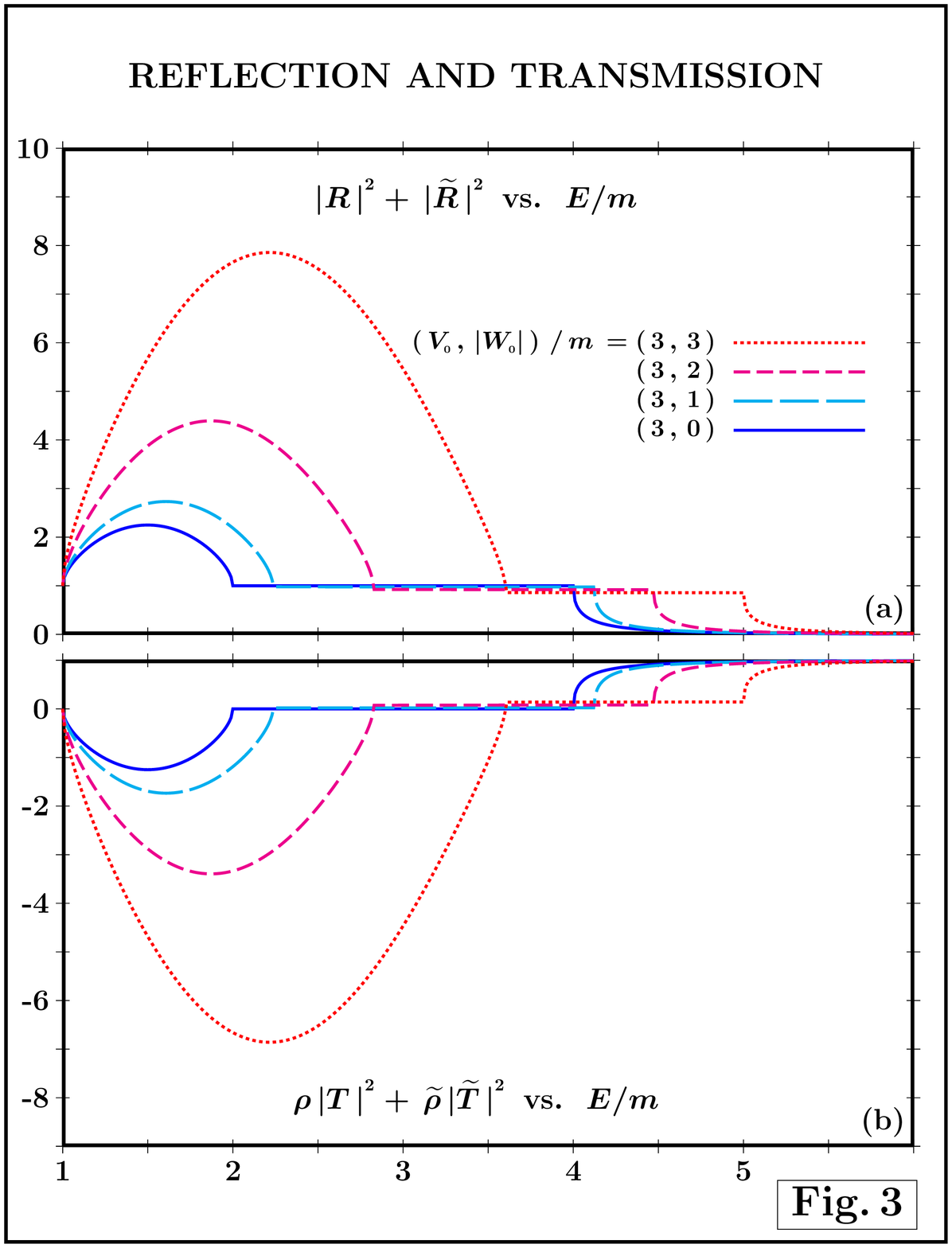}
\vspace*{-2.5cm}
 \caption{The reflection and transmission rates satisfy the probability conservation. In the Klein zone, we find a negative transmission flux and consequently a reflection rate greater than one. This effect is explaind in terms of pair production. The quaternionic perturbation amplify the pair production.}
\end{figure}

\newpage

\begin{figure}
\vspace*{-3cm} \hspace*{-2cm}
\includegraphics[width=19.5cm, height=26cm, angle=0]{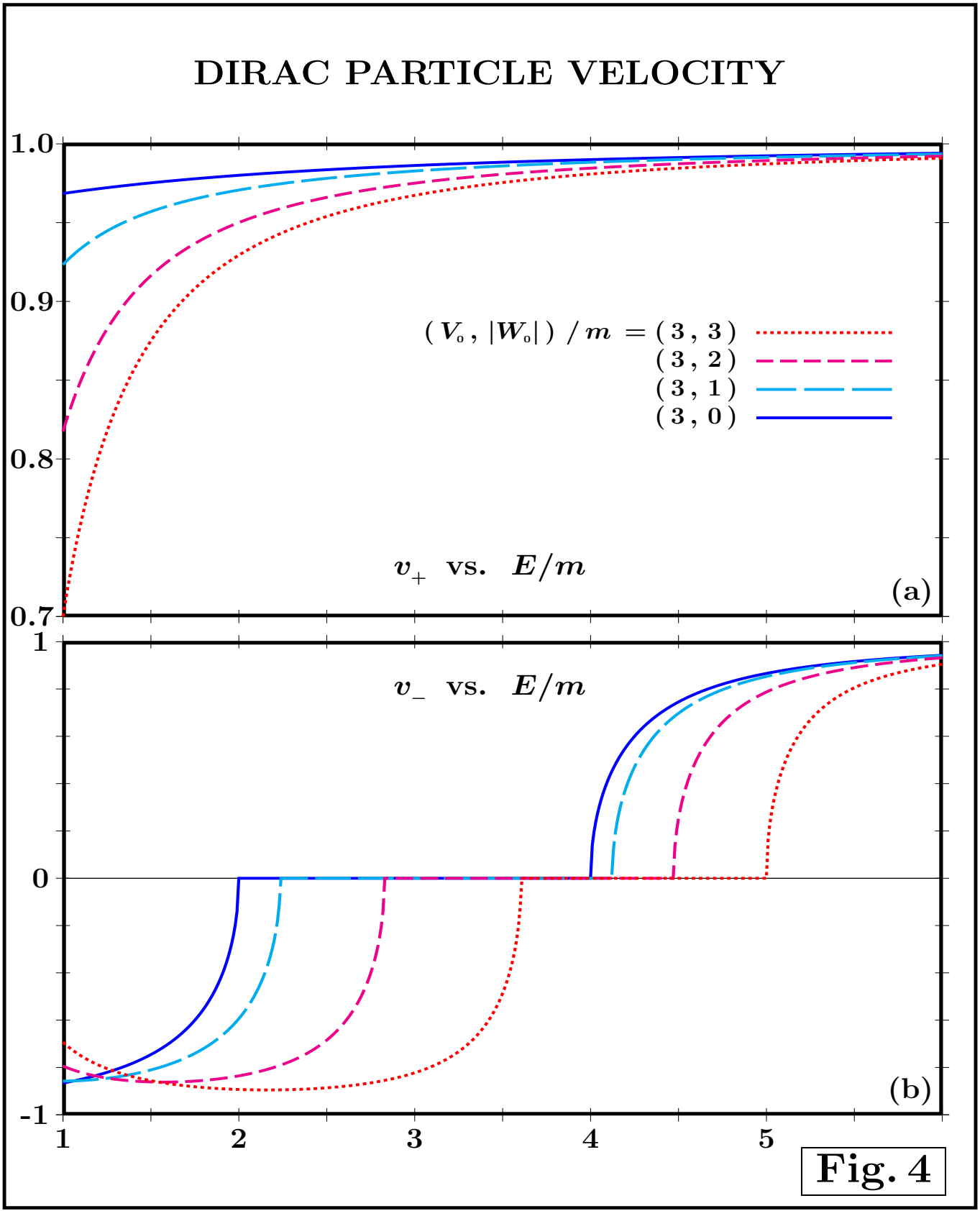}
\vspace*{-2.5cm}
 \caption{The group velocities, plotted as function of the incoming energy, confirms the analysis done for the reflection and transmission rates.  The quaternionic perturbation acts in the diffusion zone by decelerating the particle and by decreasing the tunneling energy zone.}
\end{figure}

\end{document}